\documentclass[preprint,12pt]{elsarticle}

\usepackage{amssymb}

\usepackage{amsmath}
\usepackage{caption}
\usepackage{subcaption}
\usepackage{graphicx}
\usepackage{mwe}
\usepackage{url}

\journal{Physica A}

\begin{document}

\begin{frontmatter}

%

%
%

\title{Effects of Network Structure on the Performance of a Modeled Traffic Network under Drivers' Bounded Rationality}


\author[label1]{Toru Fujino}
\ead{tr.fujino@scslab.k.u-tokyo.ac.jp}
\author[label1]{Yu Chen}
\ead{chen@k.u-tokyo.ac.jp}
\address[label1]{Graduate School of Frontier Sciences, University of Tokyo, 5-1-5, Kashiwanoha, Kashiwa, Chiba 277-8561, Japan}

\begin{abstract}
We propose a minority route choice game to investigate the effect of the network structure on traffic network performance under the assumption of drivers' bounded rationality.
We investigate ring-and-hub topologies to capture the nature of traffic networks in cities, and employ a minority game-based inductive learning process to model the characteristic behavior under the route choice scenario.
Through numerical experiments, we find that topological changes in traffic networks induce a phase transition from an uncongested phase to a congested phase.
Understanding this phase transition is helpful in planning new traffic networks.

\end{abstract}


\end{frontmatter}


\section{Introduction}
There are various types of traffic networks, e.g., road traffic networks, railway networks, and aviation networks.
These underlying infrastructures support numerous social and economic activities.
Hence, understanding the complex behavior and performance of traffic systems based on such networks has attracted considerable attention across a wide range of disciplines, including traffic engineering~\cite{Nakayama1999, Nakayama2000, Nakayama2001}, economics~\cite{Selten2007}, physics~\cite{Ezaki2015}, and computer science~\cite{Roughgarden2005}.

The main difficulty in analyzing traffic network performance lies in deviations from the user equilibrium (UE), in which drivers in traffic networks are assumed to be completely rational and have perfect information about their route choice behavior.
In real situations, the above two assumptions of UE are invalid.
It is known from the Selten's study~\cite{Selten2007} that UE can not be reached even in an idealized system, where drivers have the same origin and destination pair and only two routes to choose from.
Laboratory experiments in their study demonstrate that the number of drivers on each route fluctuates until the end of the experiment, even though the mean number of drivers on each route becomes close to the equilibrium.

Agent-based models (ABMs) provide a good framework for dealing with the bounded rationality of drivers on abstract networks.
In ABMs, drivers are represented by agents with limited intelligence.
A number of studies have considered such a framework.
Nakayama et al.~\cite{Nakayama1999} built a driver model with cognition and learning under the route-choice scenario, and they found that the network flow does not necessarily converge to UE, but may reach the deluded equilibrium, which can be derived from the drivers' false perception of the environment.
Gourley et al.~\cite{Gourley2006} used a Minority Game (MG) to formulate the agents' route choice behavior between a shortcut with variable congestion costs and a detour with a constant cost.

One factor that has a nontrivial influence on the traffic network performance is the variation in traffic network structures.
In the real world, there are numerous types of structured road networks, which makes it more difficult for drivers to select an optimal route among the alternatives.
Many studies from this viewpoint have focused on the UE solution~\cite{Wu2006, Wu2008, Youn2008, Zhao2007, Zheng2013, Zhu2014}.


These influential aspects of traffic network analysis, however, have not been combined in a unified framework, and so these analyses have not been comprehensive.
Although previous studies on ABMs have incorporated the simple aspects of the drivers’ bounded rationality, another important aspect of bounded rationality, such as preference heterogeneity, has not yet been considered, and variations in network structures have not been investigated. In the previous researches on UE, the effects of network structures have been examined comprehensively, but these studies have not dealt with the drivers' bounded rationality.

In this paper, we aim to overcome the deficiencies in existing approaches by building a new ABM.
To deal with drivers’ bounded rationality and preference heterogeneity, we borrow the strategy of MG and Market directed resource allocation game (MDRAG)~\cite{Wang2009}.
To simulate variations in network structures, we employ Gourley’s ring-and-hub topology~\cite{Gourley2006}.
In addition, we explore the optimal network structure under these conditions for implication to plan new traffic networks.


\section{Model}
\subsection{Overview of the simulation model}
The complexity of the route choice problem can be understood by the following scenario similarly explained by Gourley~\cite{Gourley2006}.
When you are the only traveler on a road traffic network, it is not difficult to determine the quickest route to get to your destination.
However, in most cases, a large number of travelers must simultaneously choose a route along a common network to reach their respective destinations.
This condition makes the traffic situation complicated.
If too many travelers select the same shorter (or quicker) route, the route tends to become extremely congested.
In such cases, making a detour may be a better choice.
At the system level, if the travelers were perfectly rational and had perfect information, it would be possible to attain the UE state, wherein no driver could reduce the transportation cost by switching to a different route.
However, as drivers are boundedly rational in reality, i.e., traffic information is incomplete and their capacity to evaluate routes is limited, it is impossible for them to make perfect decisions.
In addition, each driver has a different preference for choosing his or her route, and this heterogeneity leads to complicated situations.
Therefore, UE can rarely be attained in real situations.

To tackle this complexity and achieve the research objectives, we construct a new agent-based model, based on the Gourley's model~\cite{Gourley2006}, with the extension of consideration of network topological changes.
Additionally, we introduce preference heterogeneity~\cite{Wang2009} to Gourley's model in an attempt to capture the route choice behavior more realistically.
Thus, the proposed model consists of two parts, a network model, and an agent's decision-making model.

\subsection{Ring-and-hub network topology}
\begin{figure}[h]
    \centering
    \includegraphics[width=\textwidth]{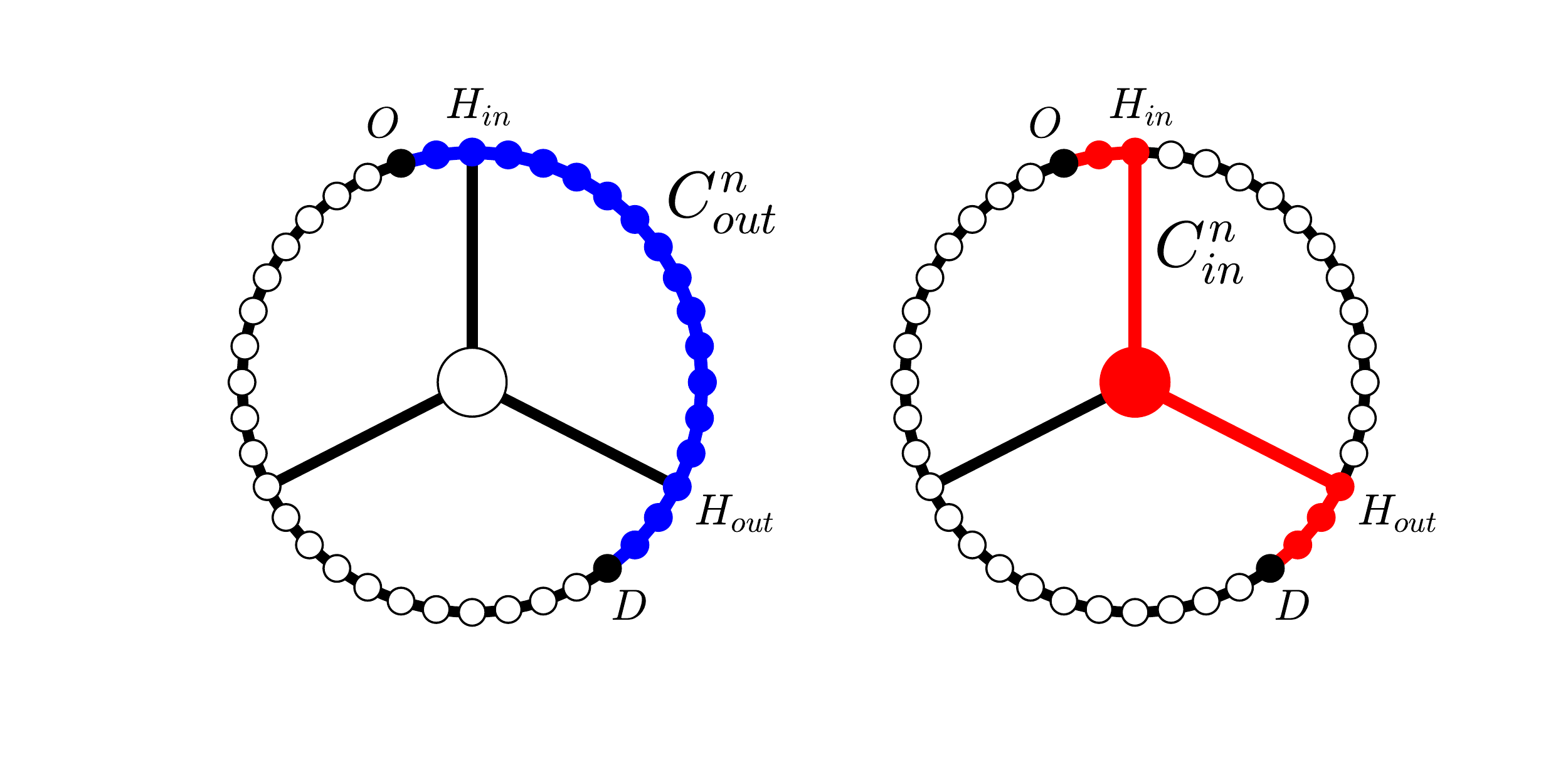}
    \caption[Ring-and-hub network]{Ring-and-hub network. The outside route is depicted by the blue line, and the inside route is depicted by the red line.}
    \label{fig:network_example}
\end{figure}

To express the route choice scenario which I described above, we employ a ring-and-hub network shown in Fig.~\ref{fig:network_example}.
This network consists of $N$ peripheral nodes, which are mutually connected by their nearest neighbors, and a central hub node, which is connected to only $\lambda \leq N$ peripheral nodes.
Each node is assigned exactly one agent, so there are $N$ agents on the network.
Each agent departs from her origin node and later arrives at her destination node, which is randomly assigned at the begining of the simulation.
There are two possible routes by which each agent can reach their destination (indicated by blue and red lines in Fig.~\ref{fig:network_example}).
One is an outside route that does not pass through the central hub, whereas the other is an inside route that passes through the central hub. The outside route can be considered as a detour, and the inside route is a shortcut.

In this modeled traffic network, the transportation costs for the outside and inside routes of agent $n=1,\dots,N$ at each time step $t=1,\dots,T$ are represented by $C^{(n)}_{out} \left(t\right)$ and $C^{(n)}_{in} \left(t\right)$, respectively.
$C^{(n)}_{out} \left(t\right)$ is simply the number of links along the shortest peripheral path connecting the origin and the destination. $C^{(n)}_{in} \left(t\right)$ depends on the number of agents in the central hub.
The cost function is given as follows:
\begin{eqnarray}
    C^{(n)}_{out} \left(t\right) & = & d_{(O, D)} \text{,}\\
    C^{(n)}_{in} \left(t\right) & = &
    \begin{cases}
        d_{(O, H_{in})} + \alpha \cdot d_{(H_{in}, H_{out})} + d_{(H_{out}, D)} \hspace{10mm} &\text{if $N_{in} \left(t\right) \leq L$} \\
        d_{(O, H_{in})} + \beta \cdot d_{(H_{in}, H_{out})} + d_{(H_{out}, D)} &\text{if $N_{in} \left(t\right) > L$.}
    \end{cases}
\end{eqnarray}
Here, $\alpha$ is a noncongestion coefficient and $\beta$ is a congestion coefficient; $L$ is the central hub capacity, which represents the limited number of agents who can be handled by the hub without congestion, and $N_{in} \left(t\right) $ is the number of central hub users at time step $t$.
$d_{(\cdot, \cdot)}$ is the length of the shortest path on the peripheral path between nodes.
$O$ and $D$ are the origin and destination nodes, respectively.
$H_{in}$ and $H_{out}$ are the interchange nodes that the agent can use to pass through the hub node along the inside route.
In this experiment, we set $\alpha$ and $\beta$ to $\frac{1}{2}$ and $\frac{3}{2}$, respectively. Thus, hub users may reach their destination at a lower cost under noncongestion, or at a higher cost under congestion.
An example is illustrated in Fig.~\ref{fig:network_example}.
In this example case, $C^{(n)}_{out} \left(t\right) $ will be $18$; $C^{(n)}_{in} \left(t\right) $ will be $11.5$ if the hub node is not congested and $24.5$ if the hub node is congested.
There may be several possible inside routes in this example, but the agents only consider the least-cost inside route among the candidates.

\begin{figure}[h]
    \centering
    \includegraphics[width=\textwidth]{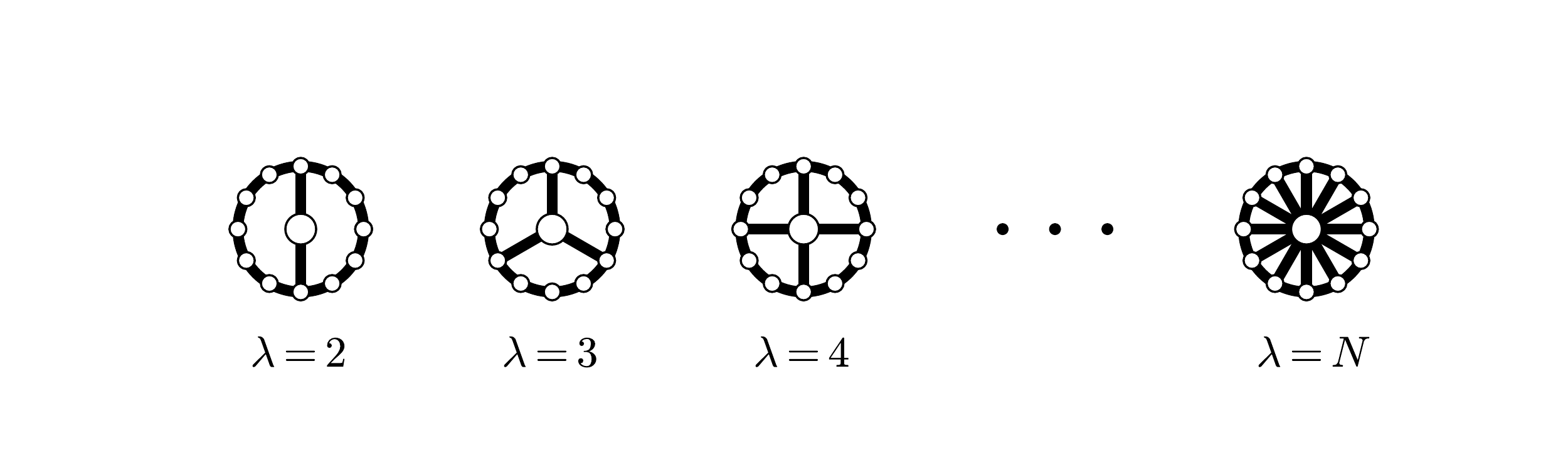}
    \caption[Ring-and-hub network]{Hub links are symmetrically added to the existing network one at a time}
    \label{fig:topological_change}
\end{figure}

To express various network structures parametrically, we vary the number of links connected to the hub node, $\lambda$, while the symmetry of the network remains unchanged, as illustrated in Fig.~\ref{fig:topological_change}.
Note that it is impractical to consider all topological combinations of ring-and-hub networks.
In addition, considering that traffic networks are artificially designed, it is not unreasonable to consider only symmetric cases.

\subsection{Minority route-choice game model}
This route choice game on the ring-and-hub network can be interpreted as a situation whereby a traveler considers whether to use a freeway to reach the destination or not.
The hub node corresponds to the freeway, and the peripheral nodes correspond to some sites with or without  an entrance of the freeway.
If the driver's origin is far from the entrance to the freeway, using a local road would be more efficient, as there would be a high cost associated with reaching a nearby interchange.
If the driver's origin is close to an interchange, he/she would use the freeway in order to get to the destination earlier.
In this case, however, the driver must also consider the possibility of traffic congestion on the freeway.

In this experiment, agents follow the Minority route choice game (MRCG), which is an inductive reasoning rule to select the routes to reach their respective destinations.
This inductive reasoning rule has a common structure with strategies in the MG case~\cite{Arthur1994, Challet1997}, because route-choice problems are very similar to the choice of minority~\cite{Chmura2006, Gourley2006}.
MRCG agents can choose from $S$ strategies.
Each strategy $s$ suggests an action $s(\mu)$ according to the past hub states in the form of an $M$-bit history string $\mu$.
At each time step $t$, the hub state $h\left(t\right)$ is recorded as either 0 (uncongested) or 1 (congested).
Hence, the total number of possible history strings is $P = 2^M$, and there are $2^P$ possible strategies.
A binary number representing an action is assigned to each history string corresponding to an agent's choice of the outside route (0) or the inside route (1).
The probability of assigning each action is determined by an integer parameter $K=0,\dots,P$~\cite{Wang2009}, which represents the tendency of the strategy to have an action $0$, where action $0$ is employed with probability $\frac{K}{P}$ and action $1$ is employed with probability $\frac{P-K}{P}$.
If $K$ is small, the strategy is likely to suggest action $0$ persistently, and vice versa.
In a homogeneous version of MRCG, we set $K = \frac{P}{2}=2^{M-1}$, where each strategy has an equal likelihood of action $0$ or $1$ for each history string.
In the heterogeneous version of MRCG, we select an integer $K$ uniformly from $0, \dots, P$ for each strategy.
An example for $M=3, K=4$ is listed in Table.~\ref{tab:mrcg_strategy}.
\begin{table}[h]
    \begin{center}
        \begin{tabular}{lcrr} \hline
            History string & Action \\ \hline
            000 & 0 \\
            001 & 1 \\
            010 & 1 \\
            011 & 1 \\
            100 & 0 \\
            101 & 1 \\
            110 & 1 \\
            111 & 0 \\ \hline
        \end{tabular}
        \caption{An example of the homogeneous MRCG strategy for $M = 3, K = 4$. In the action column, 1 denotes the inside route and 0 denotes the outside route. For instance, given the history string $010$, this strategy suggests action 1, denoting the inside route, whereas the history string $111$ leads to action 0.}
        \label{tab:mrcg_strategy}
    \end{center}
\end{table}

Among the $S$ strategies, each agent $n$ selects the strategy with the best score.
At each time step $t$, the score of each strategy $U_s\left(t\right)$ is updated based on whether the strategy achieves a lower-cost route or not.
If a strategy results in a lower cost, its score is increased.
If a strategy leads to a higher cost, its score is decreased.
Note that this update is performed regardless of whether or not the strategy is executed at that time step. The update equation is as follows:
\begin{eqnarray}
    U_s\left(t+1\right) = U_s\left(t\right) + \mathrm{sign}\left(C^{(n)}_{out}\left(t\right)-C^{(n)}_{in}\left(t\right)\right)\left(2s\left(\mu\right)-1\right)
\end{eqnarray}
where $\mathrm{sign}$ is the sign function.

\begin{figure}[h]
    \centering
    \includegraphics[width=\textwidth]{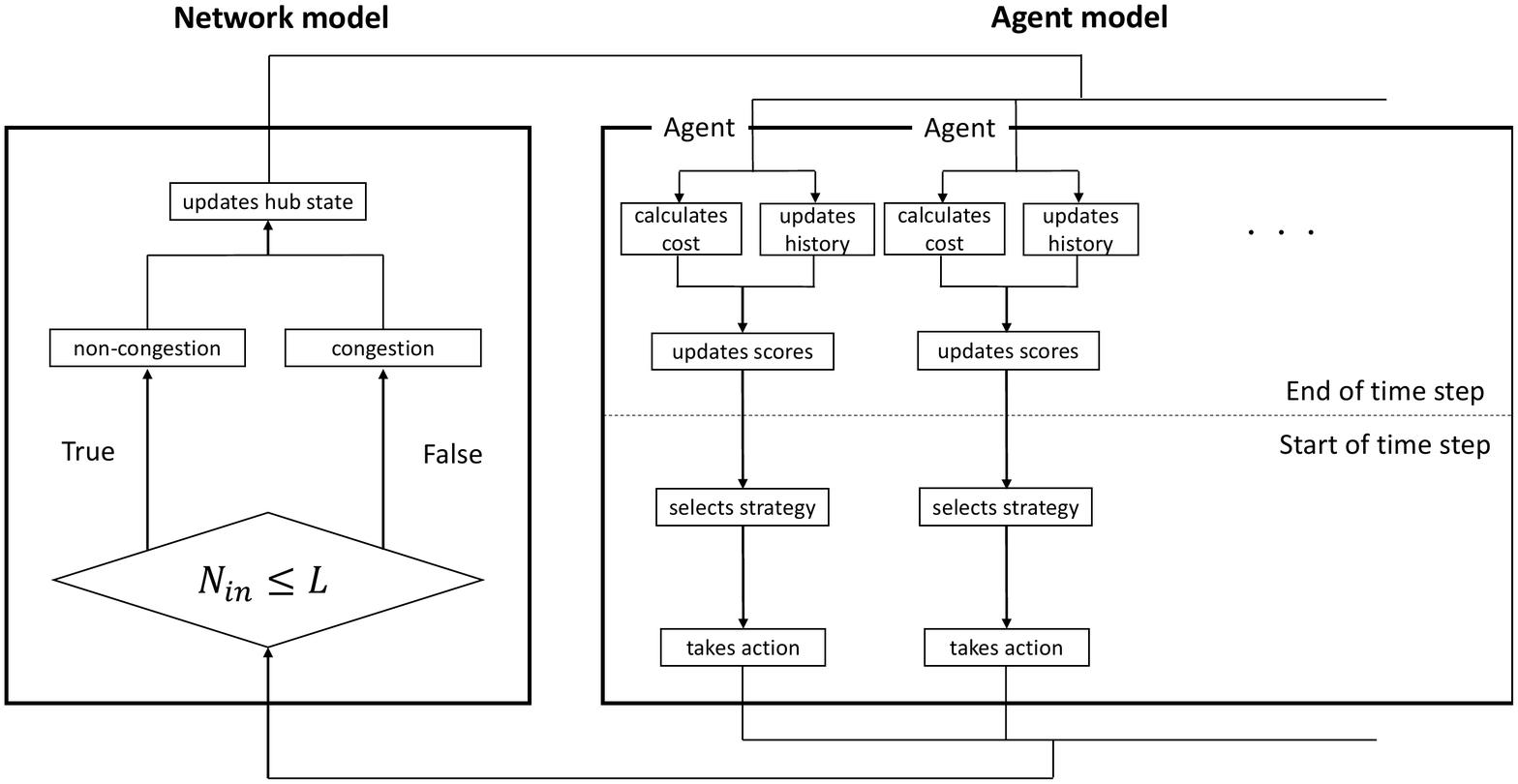}
    \caption{Simulation model outline}
    \label{fig:simulation_model}
\end{figure}

The entire structure of the simulation model is illustrated in Fig.~\ref{fig:simulation_model}.

\subsection{Evaluation of the network performance}
To assess the performance of the modeled traffic network, we employ the following basic measures.

\vspace{6mm}
\noindent
{\bf Average cost} \hspace{2mm}
The average cost is calculated by averaging costs over all agents and time steps:
\begin{eqnarray}
    \langle C \rangle = \frac{1}{TN} \sum^{T}_{t=1} \sum^{N}_{n=1} C^{(n)} \left(t\right) \text{,}
\end{eqnarray}
where $C^{(n)}(t)$ is the cost of agent $n$ at time step $t$.
Depending on the route chosen by the agent, $C^{(n)}(t)$ can be either $C^{(n)}_{out}(t)$ or $C^{(n)}_{in}(t)$.

\vspace{6mm}
\noindent
{\bf Congestion ratio} \hspace{2mm}
The congestion ratio is the ratio of the congested state to the uncongested state:
\begin{eqnarray}
    r = \frac{1}{T} \sum_{t=1} h\left(t\right) \text{.}
\end{eqnarray}

\vspace{6mm}
\noindent
{\bf Average number of hub users} \hspace{2mm}
The average number of hub users is the average number of hub users over all time steps:
\begin{eqnarray}
    N_{in} = \frac{1}{T} \sum^{T}_{t=1} N_{in}\left(t\right) \text{.}
\end{eqnarray}

\vspace{6mm}
\noindent
{\bf Standard deviation of $N_{in}\left(t\right)$} \hspace{2mm} The standard deviation of $N_{in}\left(t\right)$ is calculated over all time steps:
\begin{eqnarray}
    \sigma_{N_{in}} = \sqrt{\frac{1}{T} \sum^{T}_{t=1} \left( N_{in} - N_{in} \left(t\right)\right)^2} \text{.}
\end{eqnarray}

\vspace{6mm}
\noindent
{\bf Number of potential hub users} \hspace{2mm} Potential hub users are those agents who incur lower costs by taking the inside route in the uncongested state than by taking the outside route; i.e., $C^{(n)}_{in} \left( t | h \left(t\right) =0 \right) < C^{(n)}_{out} \left(t|h\left(t\right)=0\right)$ under noncongestion. The number of potential hub users is defined as:
\begin{eqnarray}
    N_p = \sum^{N}_{n=1} H \left( C^{(n)}_{out} \left( t | h \left( t \right) = 0 \right) - C^{(n)}_{in} \left( t | h \left( t \right) = 0 \right) \right)
\end{eqnarray}
where $H\left(\cdot\right)$ is the Heaviside step function:
\begin{eqnarray}
    H\left(x\right) =
    \begin{cases}
        0 & \text{if $x \leq 0$} \\
        1 & \text{if $x > 0$.}
    \end{cases}
\end{eqnarray}
Specifically, potential agents have incentive to use the central hub since they can reduce the cost by using it.

\section{Network structure vs network performance}
\subsection{Solution of Nash equilibrium}
To assess the effects of agents' bounded rationality, our simulation results are compared with the Nash Equilibrium (NE) solutions, as NE is easier to derive in this model than UE.
Under NE, each agent selects the optimal route, which implies that no agent can reduce the travel cost by switching route. Therefore, if $N_{p} \leq L$, all potential agents would use the hub, and if $N_p > L$, the hub will contain exactly $L$ potential agents.
Note that the hub will never be congested under NE.
Multiple solutions for NE can be derived, but for the comparison with the simulation results, we only derive the NE solution with the best allocation (the lowest cost) and that with the worst allocation (the highest cost).

When the hub node is uncongested, the difference between $C^{(n)}_{out} \left(t| h\left(t\right)=0\right) $ and $C^{(n)}_{in} \left(t| h\left(t\right)=0\right) $ is:
\begin{eqnarray}
    l^{(n)} = C^{(n)}_{out} \left(t| h\left(t\right)=0\right) - C^{(n)}_{in} \left(t| h\left(t\right)=0\right) \text{.}
\end{eqnarray}
For all agents, the differences in cost form the array
\begin{eqnarray}
    (l^{(1)}, l^{(2)}, \cdots, l^{(n)}, \cdots, l^{(N)} ) \text{,}
\end{eqnarray}
which can be sorted in ascending order as follows:
\begin{eqnarray}
    (l^{(1')}, l^{(2')}, \cdots, l^{(n')}, \cdots, l^{(N')} ) \text{.}
\end{eqnarray}
In this array, entries $1'$ to $N_p'$ are potential agents, and entries $N_p'+1$ to $N'$ are non-potential agents.
Therefore, the average cost for the NE with the best allocation, $\langle C \rangle^{NE}_b$, and that for the NE with the worst allocation, $\langle C \rangle^{NE}_w$, are:
\begin{eqnarray}
    \langle C \rangle^{NE}_b = \frac{1}{N} \Biggr( \sum_{n' \leq \min\{N_p, L \}} C^{(n')}_{in} \left(t| h\left(t\right)=0\right) + \sum_{\min\{N_p, L\} < n'} C^{(n')}_{out} \left(t| h\left(t\right)=0\right)\Biggr)
    \label{equ:best_ne}
\end{eqnarray}

\begin{eqnarray}
    \begin{split}
        \langle C \rangle^{NE}_w = \frac{1}{N} \Biggl( \sum_{n' \leq N_p - L} C^{(n')}_{out}\left(t| h\left(t\right)=0\right) &+ \sum_{N_p - L < n' \leq N_p} C^{(n')}_{in}\left(t| h\left(t\right)=0\right) \\
        &+ \sum_{N_p < n'} C^{(n')}_{out}\left(t| h\left(t\right)=0\right) \Biggr)
    \end{split}
    \label{equ:worst_ne}
\end{eqnarray}
Note that the first term in Eq.~(\ref{equ:worst_ne}) disappears when $N_p \leq L$.

\subsection{Baseline results of MRCG}
\label{sec:baseline}
\begin{figure}[h]
	\begin{minipage}{0.5\hsize}
		\centering
		\includegraphics[width=\textwidth]{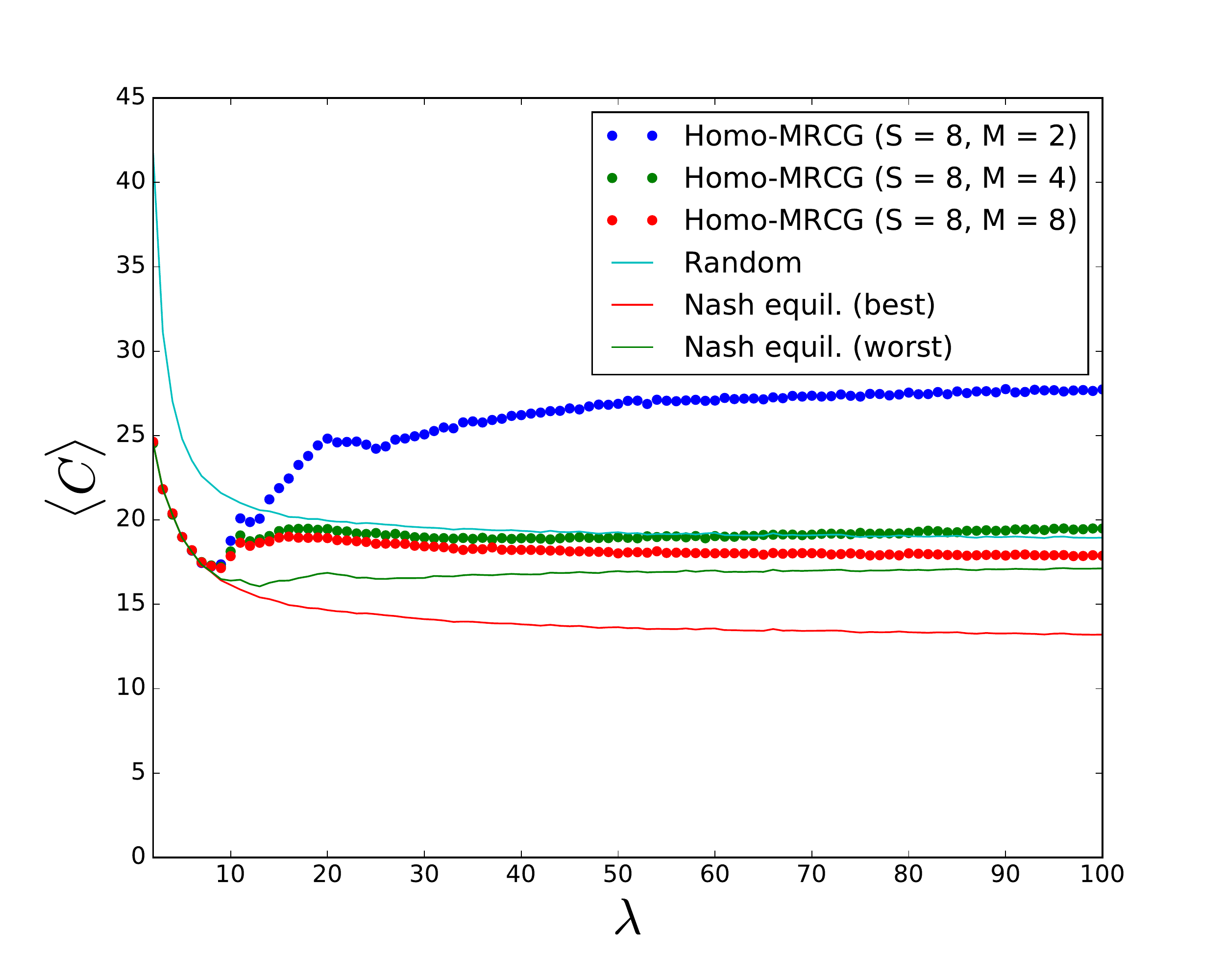}
        \caption{Average cost for homogeneous agents}
		\label{fig:c_homo}
	\end{minipage}
    \begin{minipage}{0.5\hsize}
        \centering
        \includegraphics[width=\textwidth]{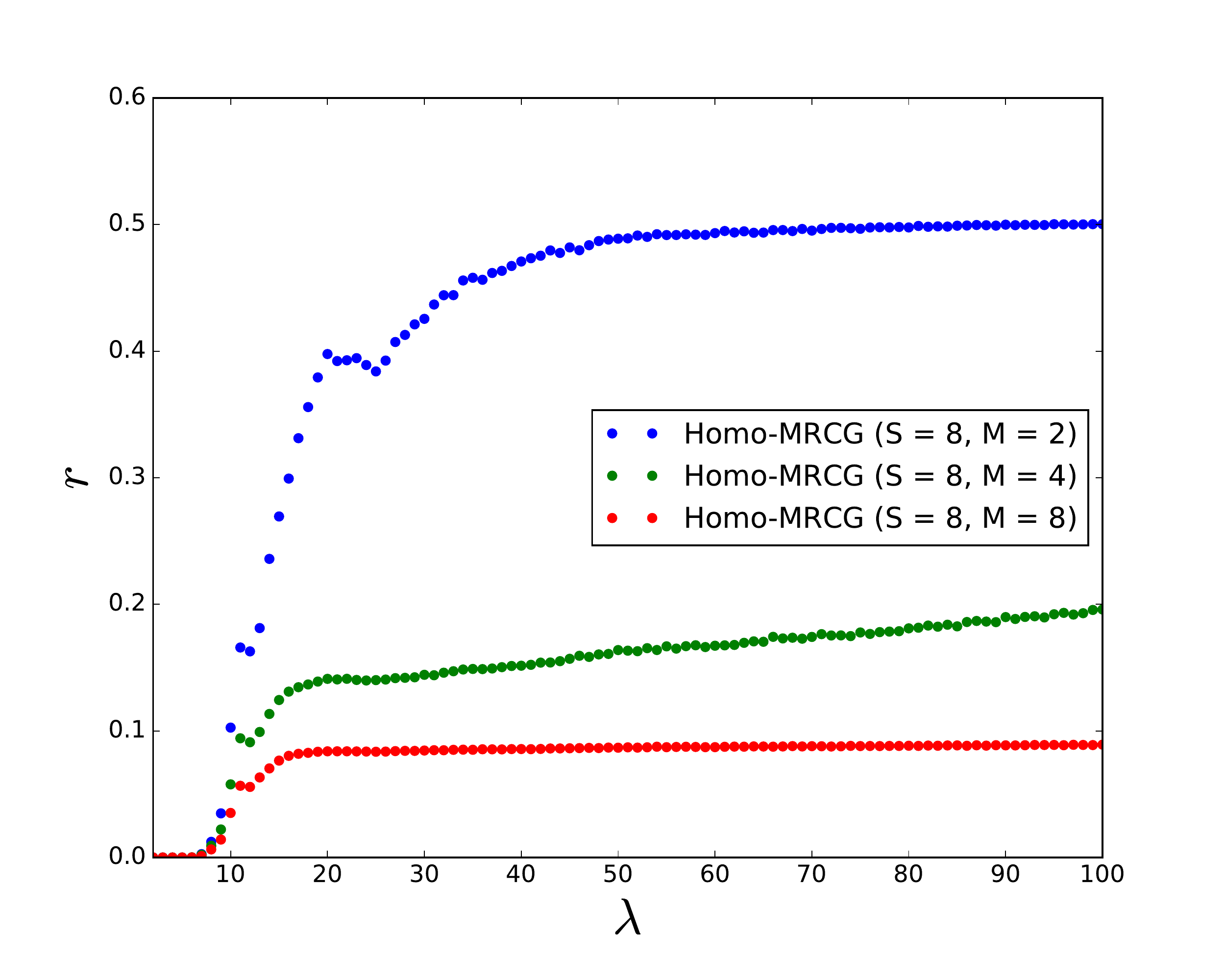}
        \caption{Congestion ratio for homogeneous agents}
        \label{fig:r_homo}
    \end{minipage}
\end{figure}
\begin{figure}[h]
	\begin{minipage}{0.5\hsize}
		\centering
		\includegraphics[width=\textwidth]{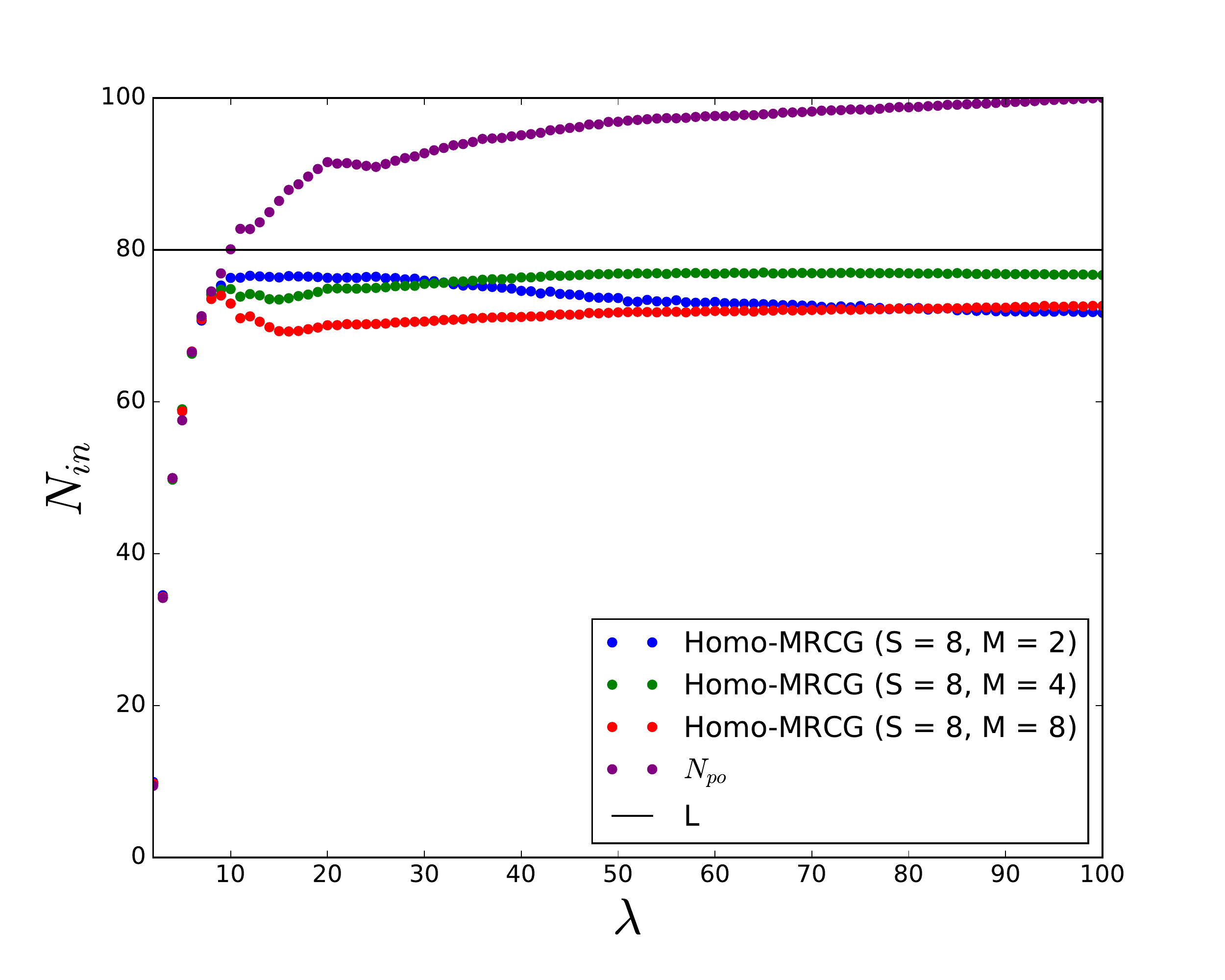}
        \caption{Number of hub users for homogeneous agents}
		\label{fig:n_homo}
	\end{minipage}
    \begin{minipage}{0.5\hsize}
        \centering
        \includegraphics[width=\textwidth]{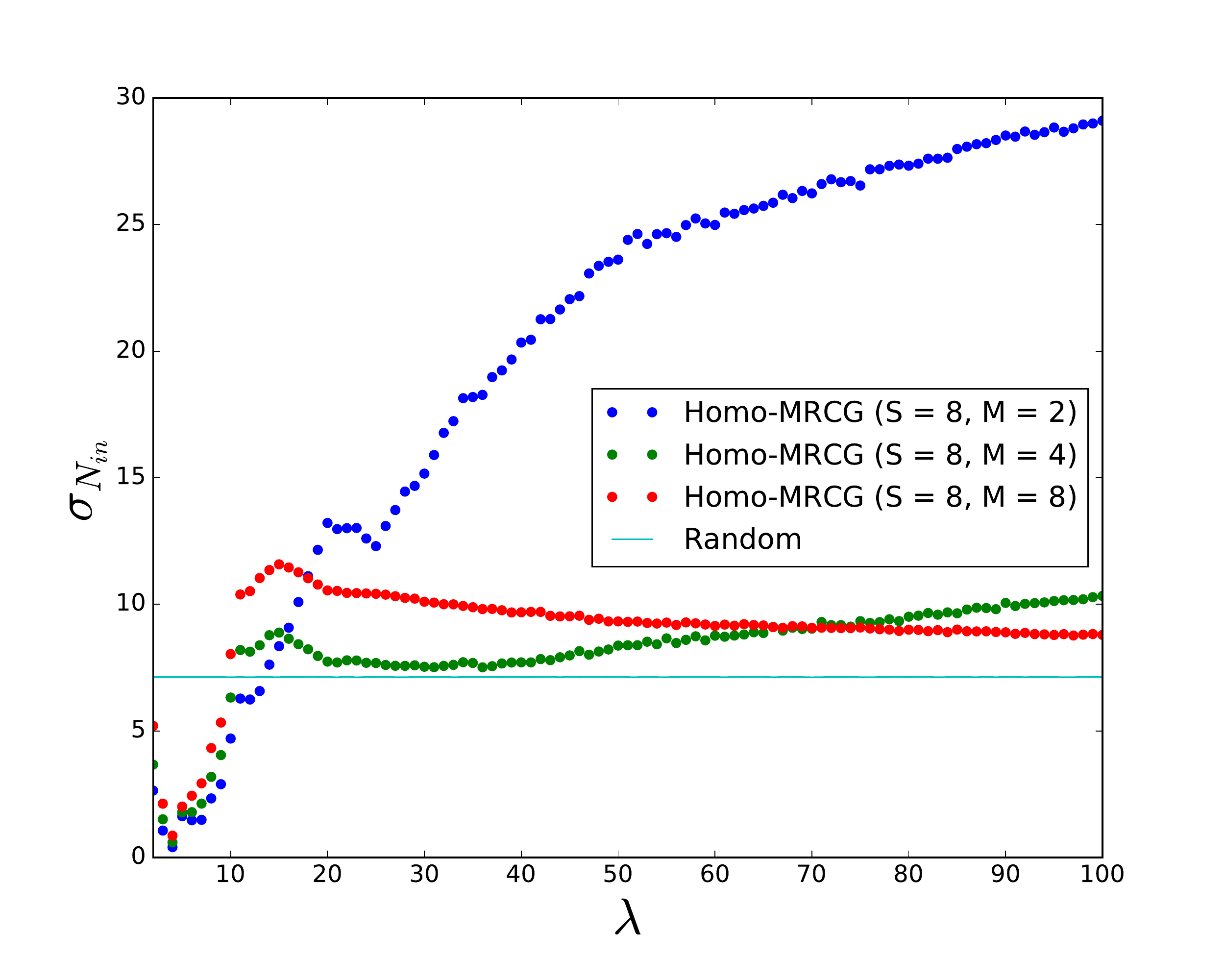}
        \caption{Standard deviation of the number of hub users for homogeneous agents}
        \label{fig:sd_homo}
    \end{minipage}
\end{figure}
For the baseline case, we set the network scale to $N = 100$ and the hub capacity to $L = 80$.
The number of hub links was varied from $2$ to $N$, and the memory length was set to $M = 2, 4, 8$.
We fixed $S = 8$ because $M$ is the dominant parameter in MG and it would be less meaningful to change $S$ and $M$ simultaneously.
Here, $K$ is fixed to $2^{M-1}$ to represent homogeneous strategies.
In each set of parameters, we computed the average results over 1000 simulations.
Each simulation was carried out for 1000 time-steps, and the first half of the simulation is for reaching the stable state, and the second half of the simulation is for getting the statistical results.
To evaluate the adaptivity of MG agents, we conducted simulations with random agents, i.e., routes were chosen at random.

From Fig.~\ref{fig:c_homo}, we can see that the simulation results behave similarly to the NE solutions in the range $2 \leq \lambda \leq 9$.
In this range, agents can adapt themselves to the environment, and the performance is much better than in the case of random agents.
For $\lambda \geq 10$, however, the homogeneous MRCG agents deviate from the NE solution.
This deviation is caused by growth in the congestion ratio at this point, as can be seen in Fig.~\ref{fig:r_homo}.
This congestion can not be seen under the assumption of the Nash equilibrium.
In the range of $\lambda \leq 9$, it is found that since the number of potential agents (purple dots in Fig.~\ref{fig:n_homo}), who can reduce their cost by using the central hub, is less than the hub capacity, the network does not suffer from congestion.
As more hub links are added to the network, $N_p$ increases, and then $N_{in} \left(t\right)$ increases as well.
This leads to the hub congestion.
We can regard this change as a phase transition from the uncongested phase to the congested phase.

It is important to investigate how different values of $M$ influence the network performances.
In the congested phase, Fig.~\ref{fig:c_homo} indicates that, as the memory length increases, the average cost decreases and becomes close to the worst case of Nash equilibrium.
In addition, $r$ decreases, enhancing the network performance.

The network stability should also be investigated in terms of the standard deviation of the number of hub users.
From Fig.~\ref{fig:sd_homo}, we can see the minimum of $\sigma_{N_{in}}$ at $\lambda = 4$ in the uncongested phase.
After this critical point, the value of $\sigma_{N_{in}}$ increases as more hub links are added to the network for $M = 2$ agents until $\lambda = N$.
In contrast, for large memory agents, $\sigma_{N_{in}}$ reaches the nearly stable state at $\lambda = 15$.
In the uncongested phase, the stability of homogeneous MRCG agents is better than that of random agents, and those with shorter memory lengths lend better stability to the system.
In this region, it can be inferred that the agents are divided into two groups, one comprising hub users and one comprising non-hub users, and $\sigma_{N_{in}}$ is better than that of random agents thanks to this division.
In the congested phase, however, the stability becomes worse than in the random agents.
The increase in memory length positively affects stability, especially in the range $\lambda \geq 70$.
Overall, in terms of stability, networks with fewer hub links perform better than those with a large number of hub links.

\begin{figure}[h]
    \centering
    \includegraphics[width=0.5\textwidth]{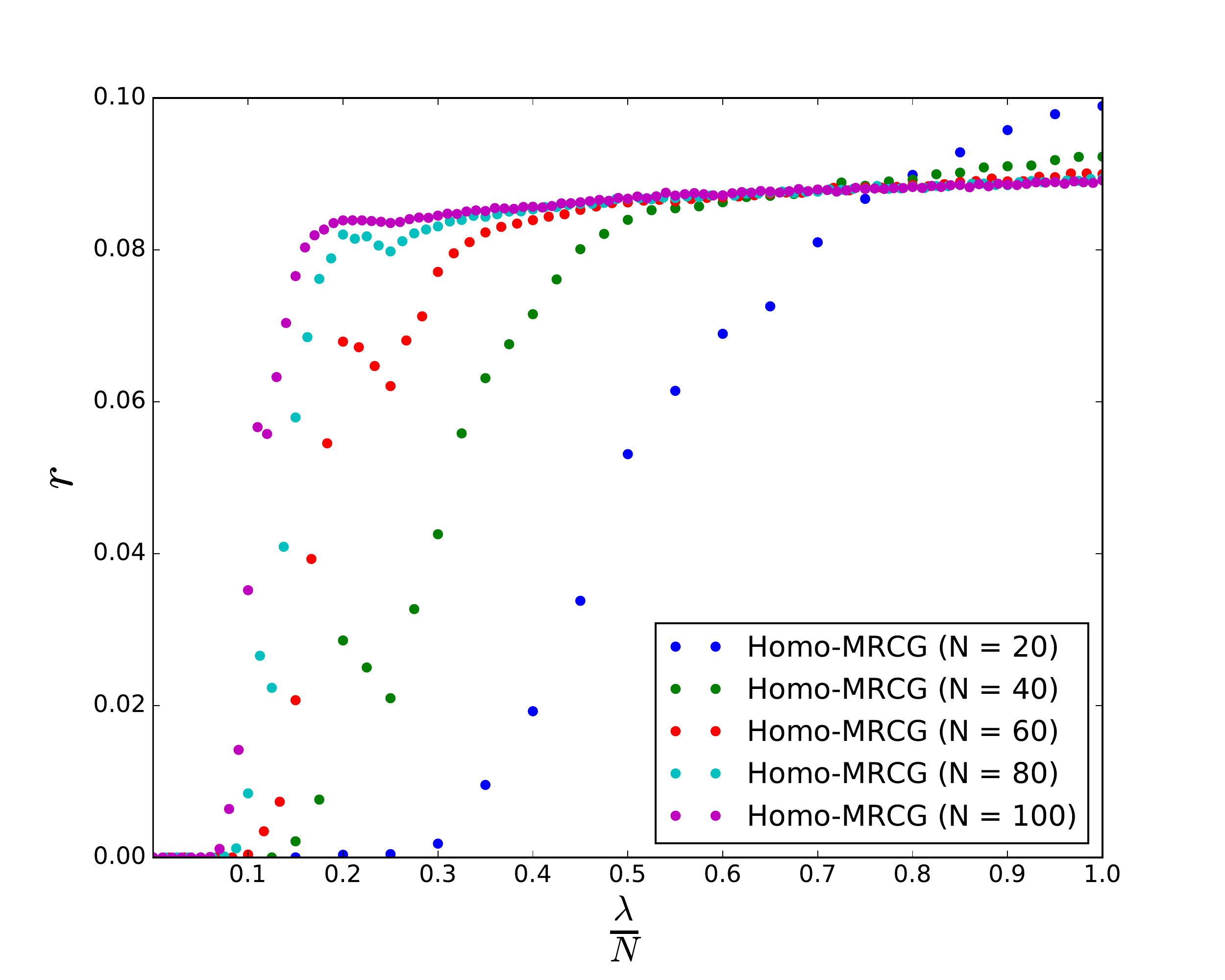}
    \caption{Congestion ratio for different scale networks}
    \label{fig:r_homo_multi}
\end{figure}
We also conducted numerical experiments for different network scales, with $N=20,40,60,80$, $S=8$, $M=2$, and $\frac{L}{N}=0.8$.
For the different networks, we can also observe a phase transition in Fig.~\ref{fig:r_homo_multi}.
Thus, this phase transition can be considered a fundamental property of agents with bounded rationality.

\subsection{Effects of preference heterogeneity on the network performance}
\begin{figure}[h]
	\begin{minipage}{0.5\hsize}
		\centering
		\includegraphics[width=\textwidth]{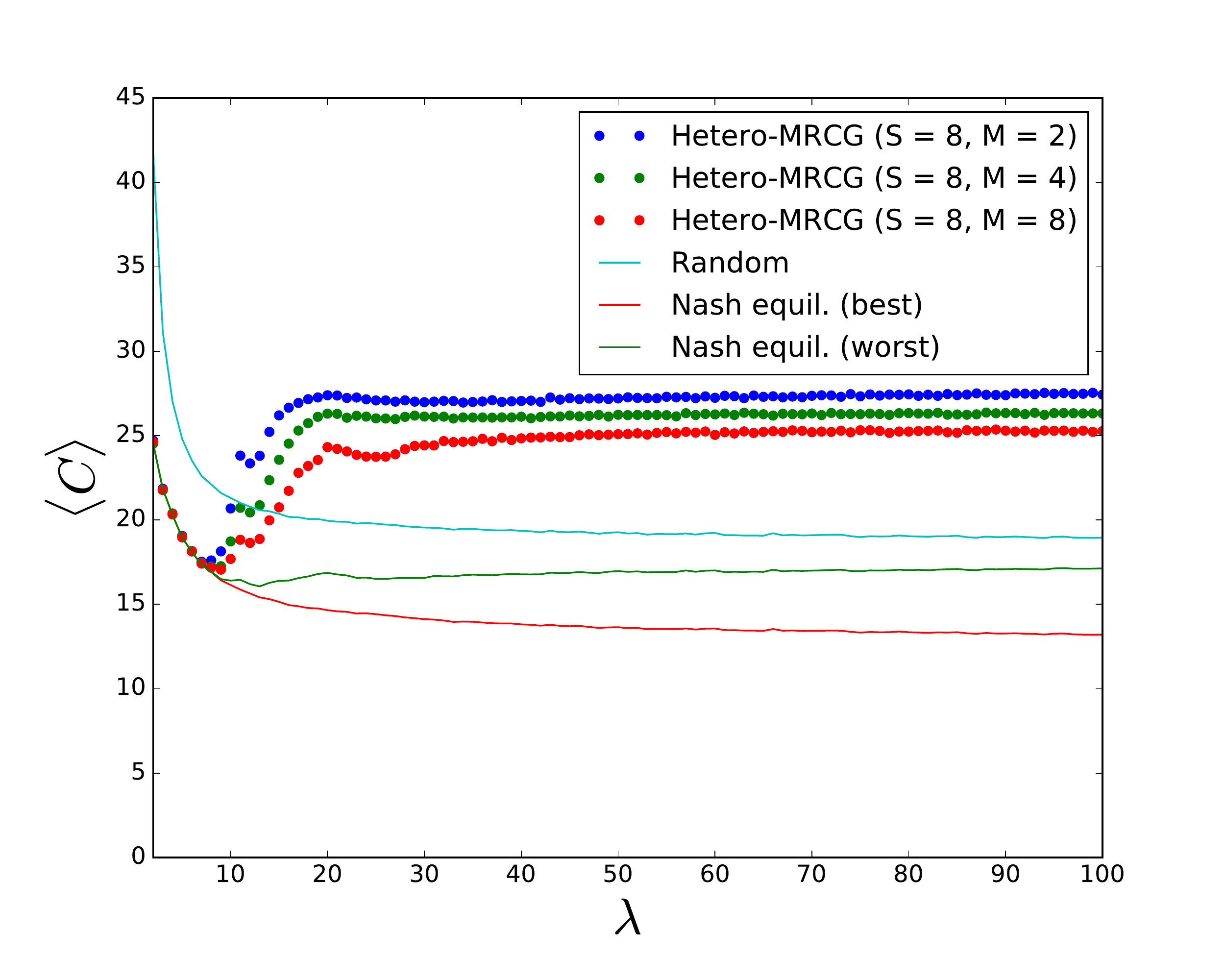}
        \caption{Average cost for heterogeneous agents}
		\label{fig:c_hetero}
	\end{minipage}
    \begin{minipage}{0.5\hsize}
        \centering
        \includegraphics[width=\textwidth]{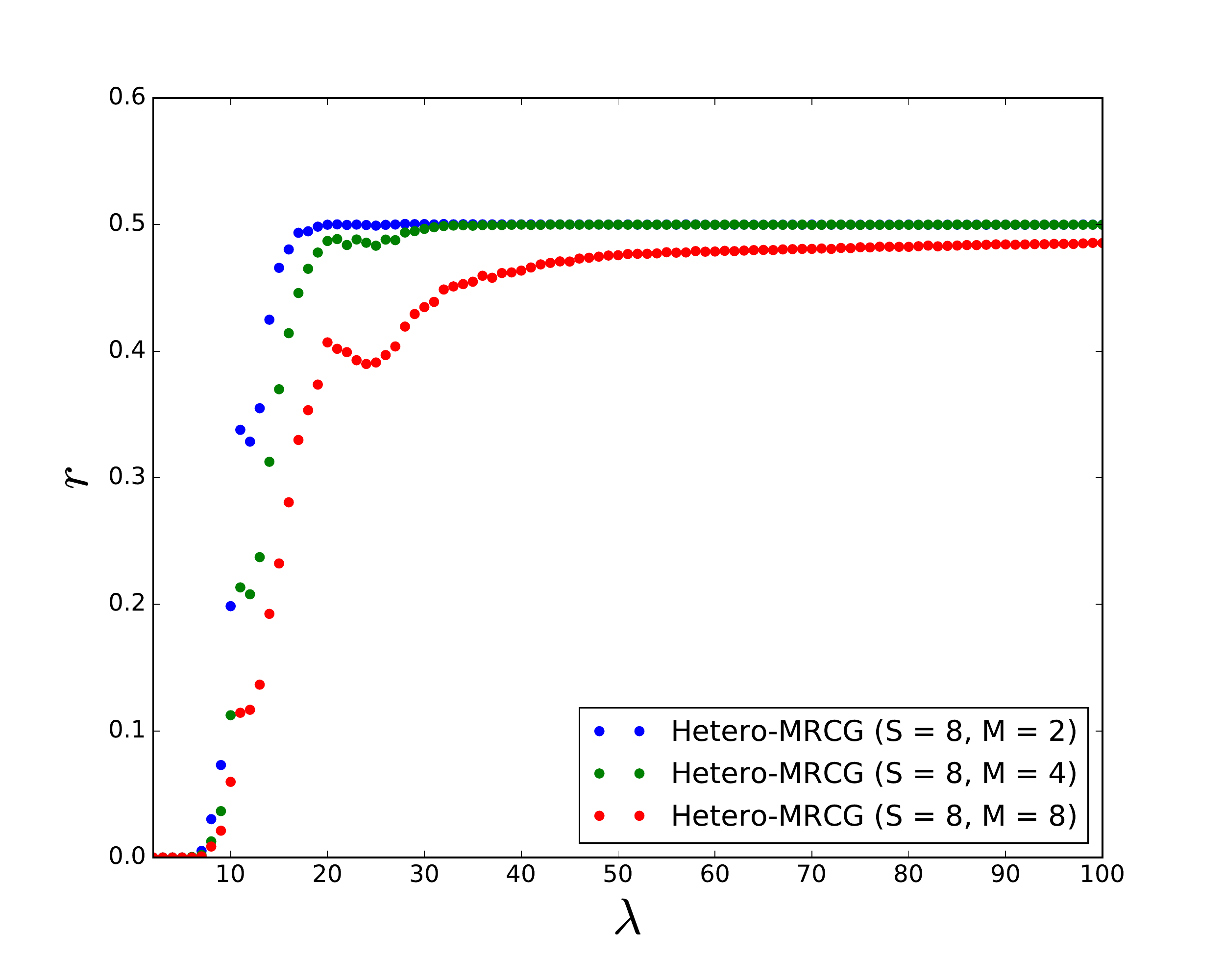}
        \caption{Congestion ratio for heterogeneous agents}
        \label{fig:r_hetero}
    \end{minipage}
\end{figure}
\begin{figure}[h]
	\begin{minipage}{0.5\hsize}
		\centering
		\includegraphics[width=\textwidth]{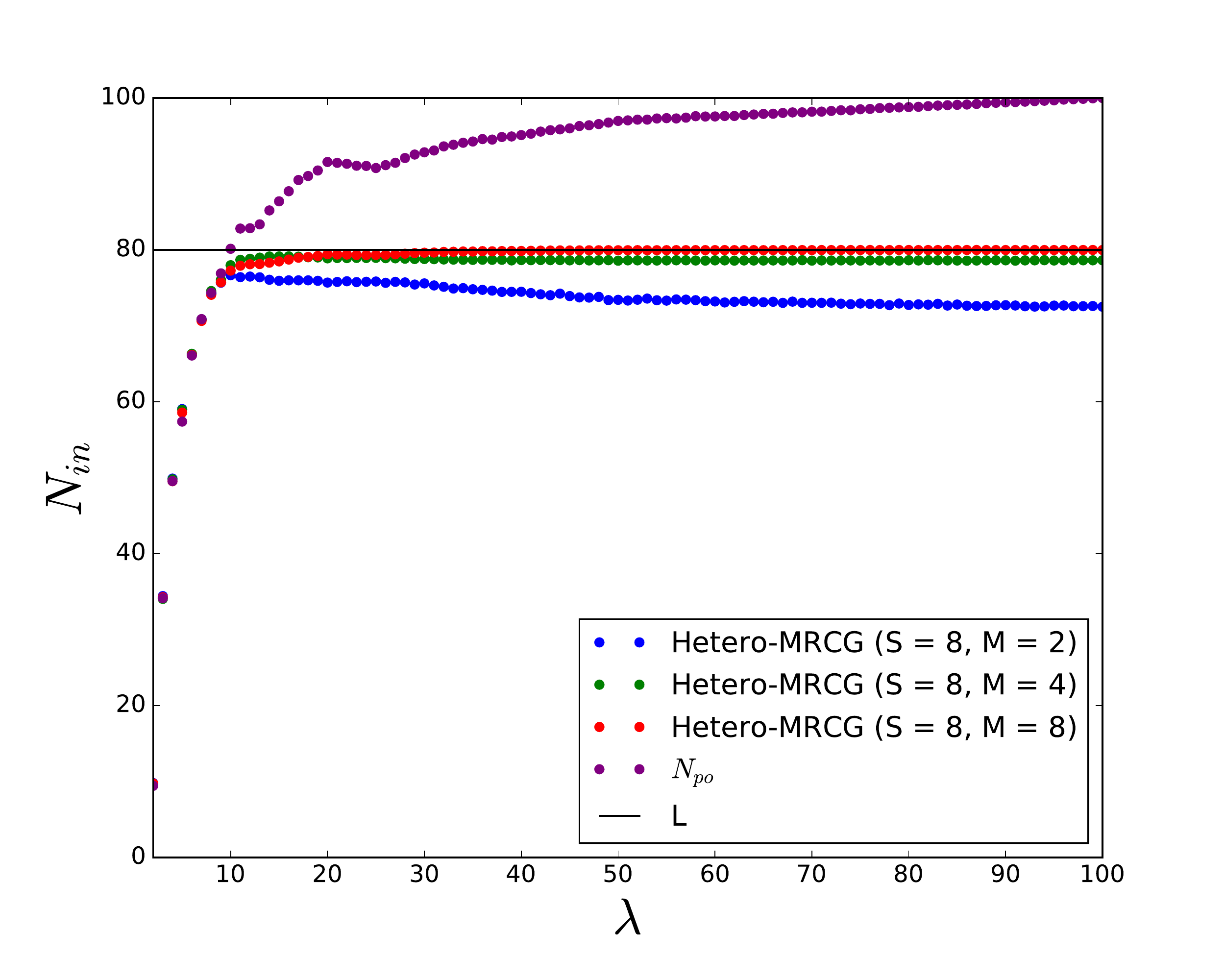}
        \caption{Number of hub users for heterogeneous agents}
		\label{fig:n_hetero}
	\end{minipage}
    \begin{minipage}{0.5\hsize}
        \centering
        \includegraphics[width=\textwidth]{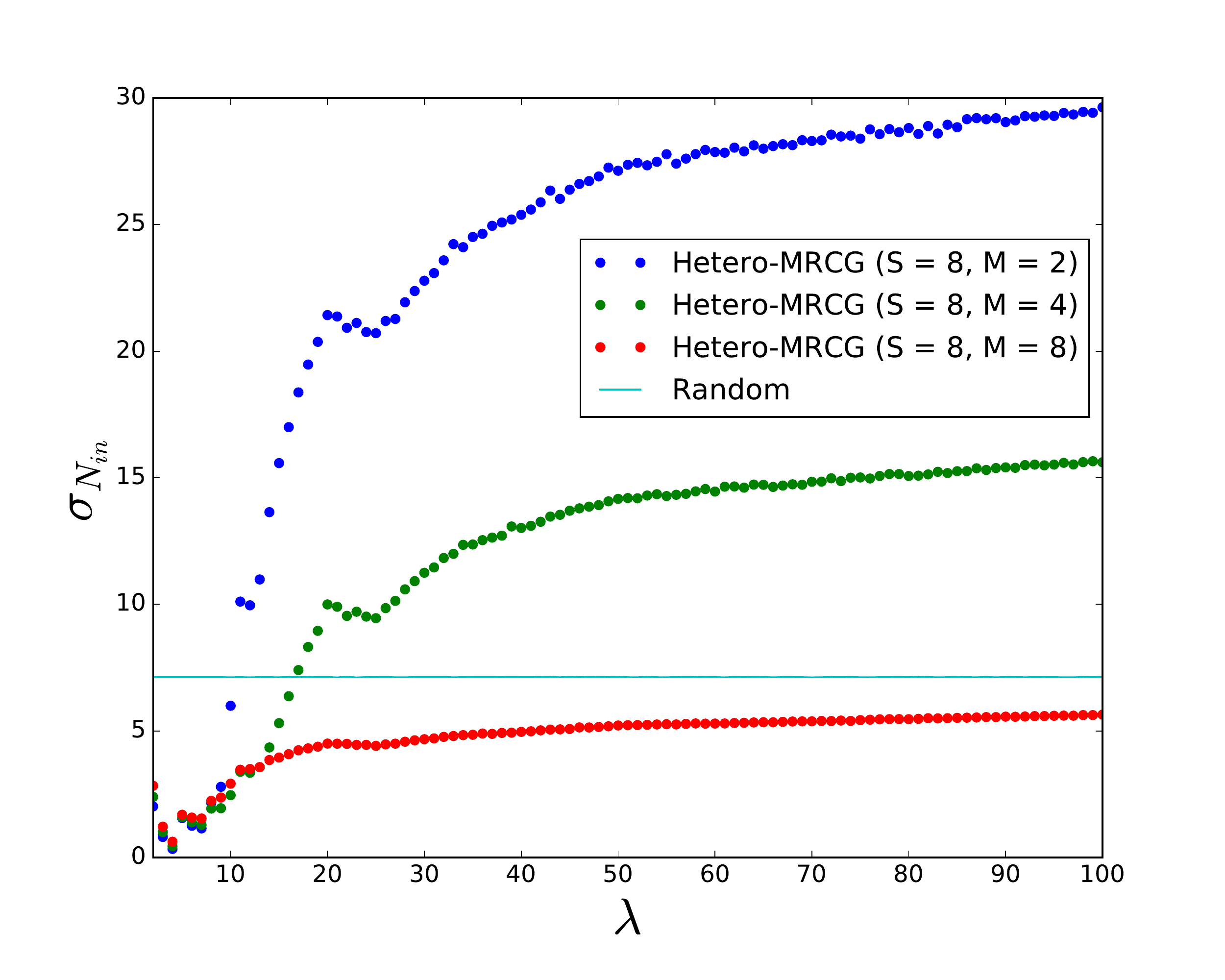}
        \caption{Standard deviation of the number of hub users for heterogeneous agents}
        \label{fig:sd_hetero}
    \end{minipage}
\end{figure}
Next, we investigated the effect of preference heterogeneity on network performance.
To this effect, we introduced preference heterogeneity into the agents’ strategies.
The value of $K$, which determines the action distribution of strategies, was set to $K = \frac{P}{2} = 2^{M-1}$ in the homogeneous case, but for the heterogeneous version of MRCG, we set $K$ by selecting integer values uniformly from $0,\dots,P$.
The other parameter settings for the simulations were the same as for homogeneous MRCG.

Similarly to the homogeneous MRCG case, we can observe a phase transition occurring at around $\lambda = 9$ in Fig.~\ref{fig:c_hetero}.
In the uncongested phase, we can see the identical behavior of $\langle C \rangle$ as discussed in \ref{sec:baseline}, giving a good match with the NE solution.
In the congested phase, the average cost becomes higher than in the homogeneous case.
The cost is also higher than in the case of random agents.
This phenomenon is because $r$ is high even for the large memory capacity (see Fig.~\ref{fig:r_hetero}), unlike the cases of homogeneous MRCG.
In contrast to the cases of homogeneous MRCG, $N_{in}$ approaches to the hub capacity as the memory length increases seen in Fig.~\ref{fig:n_hetero}.

Regarding network stability, we can also see the minimum of $\sigma_{N_{in}}$ in the uncongested phase.
As $M$ increases, the stability of the network is enhanced, as shown in Fig.~\ref{fig:sd_hetero}.
In the congested phase, similar to the homogeneous MRCG scenario, the increase in $M$ acts to reduce the value of $\sigma_{N{in}}$.
Unlike the homogeneous MRCG, however, $\sigma_{N_{in}}$ for $M=8$ agents is always below that for random agents.




\subsection{Implications of results for traffic network planning}
\begin{figure}[h]
	\centering
	\includegraphics[width=0.5\textwidth]{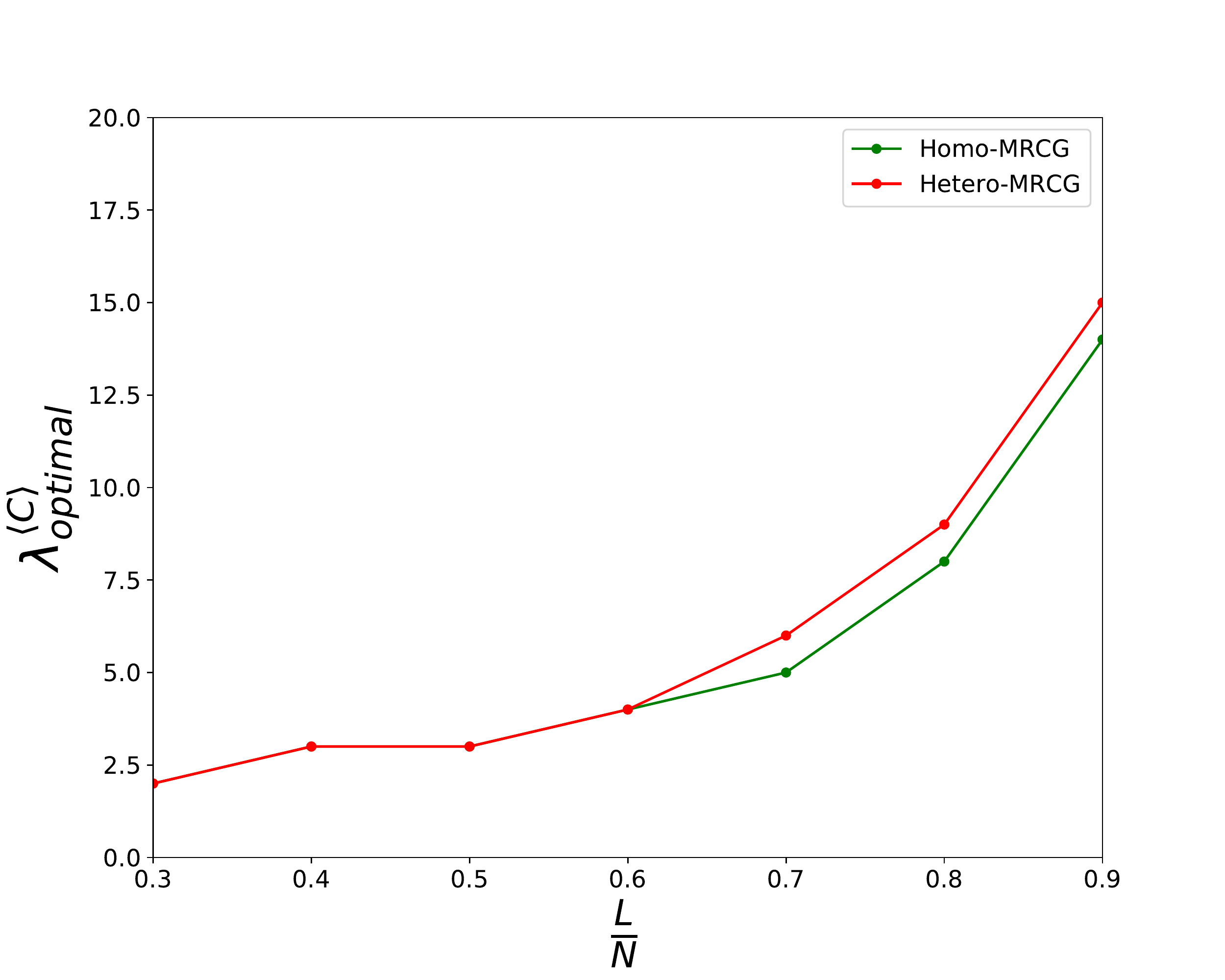}
    \caption{Optimal number of hub links}
	\label{fig:c_lmd_o}
\end{figure}
In reality, there are numerous road networks that accommodate different numbers of drivers without congestion, as they have different road capacities.
To construct well-designed networks, it is important to know the optimal network structure according to the different hub-node capacities.
For this purpose, we conducted a numerical experiment by changing the hub capacity ratio $\frac{L}{N}$ from $0.3$ to $0.9$.
We omit the results for $\frac{L}{N} = 0.2$ because the simulation result for this case does not exhibit any phase transition.
We set $N = 100, S = 8, M = 2$ for this experiment.

Fig.~\ref{fig:c_lmd_o} shows the optimal value of $\lambda$, the number of hub links that minimizes $\langle C \rangle$, for different hub capacities.
We can see that the optimal $\lambda$ increases nonlinearly as the hub capacity is augmented.
However, the optimal $\lambda$ value remains small compared to the network scale ($N = 100$), even for a large hub capacity, $\frac{L}{N} = 0.9$.
For example, when $\frac{L}{N} = 0.9$, the optimal value of $\lambda$ is less than or equal to $15$.
This suggests that traffic networks with ring-and-hub structures should have less than or equal to $15\%$ hub links.

\section{Conclusion}
In this study, we built an ABM to investigate the effects of network structures on network performance under the assumption that drivers were operating under the bounded rationality.
Through numerical experiments, we found that networks undergo a critical phase transition from the uncongested phase to the congested phase, corresponding to the particular number of hub links.
The network performance is maximized around this transition point.
While the system becomes similar to the NE case in the uncongested phase, the performance of the network in the congested phase declines and is worse than its corresponding NE performance.
In addition, we examined how agents' preference heterogeneity affects network performance.
We found that agents' preference heterogeneity enhances the system stability in terms of hub usage for large memory agents, but increases the transportation cost.
Moreover, we conducted simulations to examine the performances under the different hub node capacities.
The simulation results demonstrate that the optimal value of $\lambda$ increases nonlinearly, but remains much smaller than the whole network size $N$.

Overall, our results show the importance of considering the bounded rationality and preference heterogeneity of drivers in the modeling and research of traffic network problems, and partially answer the question of how to build a well-structured traffic network in reality.
Concretely, our simulation results imply that traffic networks with much access to a freeway may easily reach a state of congestion. In contrast, limiting access to freeways could mitigate traffic congestion and enable the Nash equilibrium to be reached.

The simulation model in this study, employing a ring-and-hub topology, focuses on a specific situation, and only symmetric network topologies have been considered.
In future works, we would like to generalize this research topic to obtain a universal perspective and apply the proposed analyses to practical problems.
In addition, a validation of the proposed model should be conducted.



\bibliographystyle{elsarticle-num}
\bibliography{ref}

%
%
%
%
\end{document}